\documentstyle[preprint,revtex,eqsecnum]{aps}
\begin{document}
\draft
\preprint{SISSA 125/92/EP}
\preprint{July, 1992}
\begin{title}
TWISTED N=2 SUPERGRAVITY\\
AS\\
TOPOLOGICAL GRAVITY IN FOUR DIMENSIONS\\
\end{title}
\author{Damiano Anselmi and Pietro Fr\`e}
\begin{instit}
SISSA-International School for Advanced Studies\\
via Beirut 2, I-34100 Trieste, Italy
\end{instit}
\begin{abstract}
We show that the BRST quantum version of pure D=4 N=2 supergravity can be
topologically twisted, to yield a formulation of topological gravity
in four dimensions. The topological BRST complex is just a rearrangement
of the old BRST complex, that partly modifies the role of physical and
ghost fields: indeed, the new ghost number turns out to be the sum of the
old ghost number plus the internal U(1) charge. Furthermore, the action
of N=2 supergravity is retrieved from topological gravity by choosing
a gauge fixing that reduces the space of physical states to the space
of gravitational instanton configurations, namely to self-dual spin
connections. The descent equations relating the topological observables
are explicitly exhibited and discussed. Ours is a first step in a programme
that aims at finding the topological sector of matter coupled N=2 supergravity,
viewed as the effective Lagrangian of type II superstrings and, as such,
already
related to 2D topological field-theories. As it stands the theory we discuss
may
prove useful in describing gravitational instantons moduli-spaces.
\end{abstract}
\eject

\section{Introduction}
\label{intro}

Recently Topological Field Theories  \cite{topolfield} have attracted a lot of
interest, both for their own sake and in connection with string theory.
Particularly interesting, because of their relation with N=2 superconformal
theories \cite{N2SCFT} and with Calabi-Yau moduli  spaces \cite{CICY}
are  topological theories in D=2 \cite{2Dtopol}.
In two dimensions the relation between N=2 supersymmetry and topological
field-theory is established via a topological twist that redefines  a new
 Lorentz group $SO(2)^{'}$ as the diagonal of the old Lorentz group with
the U(1) automorphism group of the supersymmetry algebra \cite{2Dtwist}.
In particular it implies that a whole class of N=2 correlation functions
 is topological in nature and, as such, both independent of the space-time
 points where the operators are localized and exactly calculable with
geometrical
 techniques \cite{LandauGinz}.

Notwithstanding the interest of the D=2 case, topological theories
are worth considering
 also in four-dimensions. Actually they were originally introduced in
D=4 with the discovery by Witten of topological Yang-Mills theory \cite{witten}
 and of its relation with the mathematical theory of Donaldson invariants
 \cite{donaldson} and with N=2 super Yang-Mills theory. Indeed Witten's
original
 form of Topological Yang-Mills theory, which is already gauged fixed,
was obtained via a suitable twist from D=4, N=2 Yang-Mills theory.
 The twist consists of the redefinition of a new
 Lorentz group $SO(4)^{'}=SU(2)_{L}\;\otimes\; SU(2)_{R}^{'} $
 where the factor $SU(2)_{R}^{'}$ is the diagonal of the old
 $SU(2)_{R}$ with the $SU(2)_{I}$ automorphism group of the
supersymmetry algebra.
 The general BRST-approach to this theory was developped only
later by Beaulieu and Singer
\cite{beaulieuYM}, uncovering some of the subtleties hidden in
Witten's twist approach.

{}From the experience of this example a general lesson is anyhow learnt:
just as in D=2, also in D=4 any N=2 theory is liable to a topological
twist and, as such, it should contain a topological sector where the
 correlation functions are independent of the space-time points
and exactly calculable. In particular this should apply to
 N=2 supergravity, whose topological twist must yield a gauge-fixed version of
 D=4 topological gravity. It should also apply to the hypermultiplets,
that are the D=4 counterparts of the N=2 Wess-Zumino multiplets.
 Actually, to state the conjecture in its most general form, the entire matter
 coupled N=2 supergravity, whose general form has been obtained
 in \cite{freferrara}, further generalizing the results of conformal
tensor calculus
\cite{tencalcul}, should be liable to a topological twist and have a
topological sector.

Although , the systematic programme of topologically twisting D=4,
 N=2 theories has not yet been carried  through. In this paper we try
to fill the gap
beginning with pure N=2 supergravity.

Before addressing some of the technical and conceptual details of our
derivation,
 let  us spend few words on motivations. They are essentially three:

{\sl i) The construction and the analysis of  a well founded four-dimensional
topologically gravity may furnish a gravitational analogue of Donandson theory.
In other words, it may provide a new tool to study intersection theory on
 the moduli space of gravitational instantons.}

{\sl ii) The topological interpretation should provide new calculational
tools in N=2 supergravity.}

Finally, to our taste the most exciting, although still vague motivation
is the third

{\sl iii) The special Ka\" ahler geometry \cite{SpKaler} of Calabi-Yau
moduli-space is
 related, as we already recalled, to D=2 topological field-theories.
On the other hand,
it also  follows from the requirement of N=2 supersymmetry in D=4.
{}From the superstring
 point of view, this  is understood in terms of the h-map \cite{hmap},
stating that on the
 same Calabi-Yau manifold we can compactify both the heterotic and the
type II string.
 The latter has N=2 matter coupled supergravity as an effective lagrangian.
 Hence the topological interpretation of this theory should shed new light
 on the relation between topological field-theories in two and in four
dimensions.}

Let us now outline the conceptual set up and the contents of our paper.

Our purpose is to show that the topological twist of N=2 pure supergravity
defines
a gauge-fixed version of pure topological gravity where the gauge-fixing
condition
is $\omega^{- ab}\;=\;0$, $\omega^{- ab}$ denoting the antiselfdual part
of the spin
connection.
 To this effect we utilize the BRST-approach, having, as final goal, the
comparison of
the abstract gauge-theory  a la Beaulieu-Singer \cite{beaulieu} with
the gauge-fixed approach  a la Witten \cite{witten}.

Our viewpoint on the construction of a BRST-theory is the following.
 First one singles out the classical symmetries and constructs an abstract
BRST-algebra involving only the classical fields and the ghost, with the
exclusion of
the antighosts. We name this algebra the {\sl gauge-free} BRST algebra, since,
at this level no commitment is made on the gauge fixing terms and on the
lagrangian.
Next,  in the BRST-algebra, one introduces the antighosts and the auxiliary
fields.
The choice of these latter is motivated by the gauge fixings one wants to
consider.
Finally one constructs the BRST quantum action with the given gauge-fixings.

In the case of topological gravity the gauge-free BRST algebra is the
specialization to
the Poincar\'e group of the gauge-free algebra for a topological Yang-Mills
theory.
The general form of this algebra is \cite{beaulieuYM}:
\begin{eqnarray}
sA&=&-\; \left ( {\cal D}c \; +J\; \psi \; \right ),\nonumber\\
sc&=& \phi \; - \; {{1}\over{2}} \; \left [ \; c \; , \, c \; \right ],
\nonumber\\
sF&=& {\cal D}\psi \; - \; \left [ \; c \; , \, F\; \right ],\nonumber\\
s\psi&=& {\cal D}\phi \; - \; \left [ \; c \; , \, \psi\; \right ],\nonumber\\
s\phi&=& - \; \left [ \; c \; , \, \phi\; \right ],
\label{unouno}
\end{eqnarray}
where $A=A_{\mu}  dx^{\mu}$ is the classical 1-form gauge-field, $c$
are the 0-form ghosts
(corresponding to ordinary gauge transformations
$\delta A_{\mu} \; = \; {\cal D}_{\mu} \varepsilon$ ), $\psi=\psi_{\mu}
dx^{\mu}$
is the 1-form ghost associated with the topological
symmetry ($\delta A_{\mu} \; = \; u_{\mu} $) and $\phi$ is the
0-form ghost for ghosts that has ghost number $g=2$, while the previous ghosts
have $g=1$. All fields are Lie algebra-valued.
The BRST operation $s$ in (\ref{unouno}) is manifestly nilpotent ($s^2=0$) and
anticommutes with the exterior derivative ($sd+ds=0$).

One important  ingredient of our discussion  will be
the relation between the  gauge-free BRST algebra for the ordinary
theory and for
the topological theory. It can be understood in general terms as it follows.
As it is more explicitly discussed in section II, one can extend the concept of
differential forms to that of {\sl ghost-forms}, by setting
\begin{equation}
{\hat A}~=~A \; + \; c,
\label{unodue}
\end{equation}
where $A$ and $c$ are the (1,0)
and (0,1) parts of ${\hat A}$ (a generic object of form degree $f$ and
ghost number $g$ will be described by $(f,g)$).
One can also extend the concept of exterior differentiation defining
\begin{equation}
{\hat d}~=~d \; + \; s,
\label{unotre}
\end{equation}
where $d$ and $s$ are the (1,0) and (0,1) parts of ${\hat d}$.
With these notations one finds that, expanding the extended field-strenght
\begin{equation}
{\hat F}~=~{\hat d} {\hat A} \; + \; {{1}\over{2}} \; \left [ \; {\hat A}
\; ,\; {\hat A} \; \right ]
\label{unoquattro}
\end{equation}
in its $(f,g)$ sectors, the following identifications are possible:
${\hat F}_{(2,0)}=F$, ${\hat F}_{(1,1)}=\psi$ and ${\hat F}_{(0,2)}=\phi$.
Indeed the first two equations in (\ref{unouno}) amount precisely to
these identifications,
 while the last three are sectors of the extended Bianchi identity
\begin{equation}
{\hat d}{\hat F} \; + \; \left [ \; {\hat A} \; , \; {\hat F} \; \right ] ~=~0.
\label{unocinque}
\end{equation}
Hence the gauge-free topological  BRST algebra corresponds to a
parametrization of
the extended curvature ${\hat F} $ where no constraints are imposed on
the extra components ${\hat F}_{(1,1)}$
and ${\hat F}_{(0,2)}$.

On the other hand the ordinary gauge-free BRST algebra
\begin{eqnarray}
sA&=&-\; {\cal D}c, \nonumber\\
sc&=&  - \; {{1}\over{2}} \; \left [ \; c \; , \, c \; \right ],\nonumber\\
sF&=&  - \; \left [ \; c \; , \, F\; \right ],
\label{unosei}
\end{eqnarray}
correspond to imposing the horizontality conditions ${\hat F}_{(1,1)}=0$
and ${\hat F}_{(0,2)}=0$.

This is interpreted in the framework of rheonomy \cite{fre} as follows:
 the topological BRST algebra is the {\sl off-shell} solution of
the extended Bianchi
identity (\ref{unocinque}) where all the outer components are kept
on equal footing
with the inner ones.
The ordinary gauge-free BRST algebra is instead provided by the
{\sl quantum rheonomic}
solution of the extended Bianchi identity (\ref{unocinque}) . By definition
the {\sl quantum rheonomic parametrization} is obtained from the
{\sl classical rheonomic} parametrization by replacing the
classical cotangent basis of
differential forms with  the corresponding extended one. In
this way the components
of the extended curvatures in the extended basis are the same
as the components of
the classical curvatures in the classical basis. For instance in the case of
Yang-Mills theory the classical basis is given by $A$ and $V^{a}$,
the last being the vierbein; the classical rheonomic parametrization is:
\begin{equation}
F~=~F_{ab} \; V^{a} \wedge V^{b},
\label{unosette}
\end{equation}
so that the quantum rheonomic parametrization is
\begin{equation}
{\hat F}~=~F_{ab} \; {\hat V}^{a} \wedge {\hat V}^{b}~
=~F_{ab} \; V^{a} \wedge V^{b}.
\label{unootto}
\end{equation}
Indeed, in this  case ${\hat V}^{a}= V^{a}$, the ghost part being
attached only to the
gauge field $A$, according to (\ref{unodue}), since only the gauge
transformations
are symmetries, not the diffeomorphisms.

In the case of pure gravity the classical curvatures are
\cite{fre}
\begin{eqnarray}
R^{a}&=&{\cal D}V^a
=dV^a-{\omega^a}_b\wedge V^b,\nonumber\\
R^{ab}&=&d \omega^{ab}-{\omega^a}_c \wedge {\omega}^{cb}.
\label{curvtop}
\end{eqnarray}
Their classical rheonomic parametrization is
\begin{eqnarray}
R^{a}&=0,\nonumber\\
R^{ab}&=&R^{ab}_{cd} \; V^c \wedge V^d,
\label{unodieci}
\end{eqnarray}
so that the corresponding quantum rheonomic parametrization is
\begin{eqnarray}
{\hat R}^{a}&=0,\nonumber\\
{\hat R}^{ab}&=&R^{ab}_{cd}\; {\hat V}^c \wedge {\hat V}^d.
\label{unoundici}
\end{eqnarray}
This time the vielbein being quantum extended
\begin{equation}
{\hat V}^{a} ~=~V^{a} \; + \; \varepsilon^{a}.
\label{unododici}
\end{equation}
Eq.s (\ref{unoundici}) lead to the BRST algebra associated with diffeomorphisms
and Lorentz rotations.
On the other hand, if we relax (\ref{unoundici}) and we keep all the
outer components
 of ${\hat R}^{ab}$ as independent fields,  we obtain a gauge-free
BRST algebra that
 includes also the ghosts for the topological symmetry
$\delta V^{a}_{\mu}~= ~\xi^{a}_{\mu} $,  $\xi^{a}_{\mu} $ being an arbitrary
 infinitesimal vierbein. This is our definition of gravitational
topological BRST algebra.
 We want to compare it with the BRST algebra associated with twisted
N=2 supergravity.
  Indeed in order to make a successful twist we must already start at the
  quantum BRST-level.

Our logical development is the following.

In section II we consider N=2 supergravity in the rheonomy framework
and we construct
its BRST quantization. In particular we discuss its gauge-free BRST
algebra prior to
the introduction of antighosts.

In section III we define D=4 topological gravity along the lines
discussed above and we
 introduce the gauge-free topological BRST algebra. We also discuss the descent
 equations arising from the topological observables associated with
the Pontriagin
 and Euler characteristic classes.

In sections IV and V we discuss the topological twist of N=2 supergravity,
introducing also
the concept of {\sl topological shift} that is instrumental for a correct
 interpretation of the resulting theory. We identify the ghosts and
antighosts and from
the latter identification we conclude that the gauge-fixing implicit in the
 theory is $\omega^{- ab}\;=\;0$.

In section VI we show that the action of N=2 supergravity can be obtained as
 a topological term plus the BRST-variation of a gauge fermion $\Psi$
that implements
the gauge-fixing $\omega^{- ab}\;=\;0$. Some subtleties related with
the redundancy
of this gauge-fixing and with the appearance of extraghosts are also discussed.

Finally section VII contains our conclusions.

\section{BRST-quantum version of D=4 N=2 supergravity}
\label{brsn=2}

In this section we construct the BRST quantization of D=4
N=2 supergravity. As anticipated in section I, we think it
convenient to employ the
formalism of differential forms and rheonomic parametrizations
\cite{fre}. The concepts of rheonomy is applied to the construction
of the BRST-quantum version of the theory in the way
explained in Ref.\ \cite{beaulieu}.

D=4 N=2 simple supergravity is described by the following curvatures
\cite{fre}
\begin{eqnarray}
R^a&=&{\cal D}V^a-{i\over 2}\bar\psi_A\wedge\gamma^a\psi_A
=dV^a-{\omega^a}_b\wedge V^b-{i\over 2}\bar\psi_A\wedge\gamma^a\psi_A,
\nonumber\\
R^{ab}&=&d \omega^{ab}-{\omega^a}_c \wedge {\omega}^{cb},\nonumber\\
\rho_A&=&{\cal D}\psi_A=d\psi_A-{1\over 2}\omega^{ab}\wedge\sigma_{ab}\psi_A,
\nonumber\\
R^{\otimes}&=&F+\epsilon_{AB}\bar\psi_A\wedge\psi_B,
\label{curvatures}
\end{eqnarray}
where Lorentz indices are denoted by latin letters,
$V^a$ is the one form representing the vierbein, $\omega^{ab}$
the one form representing the spin connection, $\psi_A$ (A=1,2) is the couple
of gravitinos (one forms as well), while $F\equiv dA$, $A$ being the
one form representing the graviphoton. $d$ denotes the operation of exterior
derivative, while ${\cal D}$ represents the covariant exterior derivative.
Finally, $\sigma^{ab}\equiv {1\over 4}[\gamma^a,\gamma^b]$ and $\epsilon_{AB}$
is the completely antisymmetric tensor with two indices.

The above curvatures satisfy the following Bianchi identitites \cite{fre}
\begin{eqnarray}
{\cal D}&&\!\!R^a+{R^a}_b \wedge V^b -i\bar\psi_A\wedge \gamma^a \rho_A=0,
\nonumber\\
{\cal D}&&\!\!R^{ab}=0,\nonumber\\
{\cal D}&&\!\!\rho_A+{1\over 2}R^{ab}\wedge \sigma_{ab}\psi_A=0,\nonumber\\
{\cal D}&&\!\!R^{\otimes}+2\epsilon_{AB}\bar\psi_A\wedge\rho_B=0.
\label{bianchi}
\end{eqnarray}

The rheonomic parametrizations of the four curvatures (\ref{curvatures})
that are compatible with the Bianchi identities (\ref{bianchi}),
at least on shell, since we do not introduce auxiliary fields, are
\cite{fre}
\begin{eqnarray}
R^a&=&0,\nonumber\\
R^{ab}&=&{R^{ab}}_{cd} V^c\wedge V^d+\bar \theta^{ab}_{A|c} \psi_A\wedge V^c-
{1\over 2}\bar \psi_A\wedge {\cal F}^{ab}\psi_B\epsilon_{AB}, \nonumber \\
\rho_A&=&\rho_A^{ab}V_a\wedge V_b+{1\over2}i \gamma^a{\cal F}_{ab}
\psi_B\wedge V^b\epsilon_{AB}, \nonumber\\
R^{\otimes}&=&F_{ab}V^a\wedge V^b,
\label{rheo}
\end{eqnarray}
where ${\cal F}^{ab}\equiv F^{ab}+{i\over 2}\gamma_5 F_{cd}\varepsilon^{abcd}$
and $\bar\theta^{ab|c}_{A}=2i\bar\rho_A^{c[a}\gamma^{b]}-
i\bar\rho_A^{ab}\gamma^{c}$
\cite{fre}, where the square brakets denote antisymmetrization.
These parametrizations are found by expanding the curvatures (\ref{curvatures})
in a basis
of differential forms in superspace (which can be written as
exterior products of $V^a$ and $\psi_A$)
and then imposing the Bianchi identities (\ref{bianchi}) on shell \cite{fre}.

For completeness, we write here the lagrangian of N=2 supergravity, because it
will be useful later on.
\begin{eqnarray}
{\cal L}&=&R^{ab}\wedge V^c\wedge V^d\varepsilon_{abcd}+4\bar \rho_A\wedge
\gamma_5\gamma_a
\psi_A\wedge V^a+2iR^{\otimes}\wedge\bar\psi_A\wedge\gamma_5\psi_B
\epsilon_{AB}+\nonumber\\
&-&2i\bar\psi_A\wedge\psi_B\wedge\bar\psi_A\wedge\gamma_5\psi_B
-F^{ab}V^c\wedge V^d\wedge R^{\otimes}\epsilon_{abcd}+
\nonumber\\&+&{1\over 12}
F_{ab}F^{ab}V^i\wedge V^j\wedge  V^k\wedge  V^l\varepsilon_{ijkl}.
\label{lagra}
\end{eqnarray}

The operation of BRST transformation is denoted by $s$. We introduce
ghost-number and $s$ has ghost-number one. In such a way we have two natural
gradations: form-number $f$ and ghost-number $g$. As anticipated in the
introduction, a generic object is described
by the couple $(f,g)$. When permuting two objects it is the sum $f+g$ that
determines the correct sign (but note that some fields, like the gravitinos and
their ghosts also have a fermionic number and when permuting two of them, the
preceding rule must by suitably amended). So, $f+g$ is a gradation of
primary importance.
We shall call it the {\sl ghost-form-number}. Any object must have
a well defined
ghost-form-number and so the first part of BRST quantization consists in
extending any differential form (of form-number $f$, say) to a {\sl ghost-form}
of ghost-form-number $f$. Let
\begin{eqnarray}
\hat V^a&=&V^a+\varepsilon^a,\nonumber\\
\hat\omega^{ab}&=&\omega^{ab}+\varepsilon^{ab},\nonumber\\
\hat\psi_A&=&\psi_A+c_A,\nonumber\\
\hat A&=&A+c,
\label{hat}
\end{eqnarray}
where $\varepsilon^a$, $\varepsilon^{ab}$, $c_A$ and $c$ (form-number zero,
ghost-number one) are the ghosts of diffeomorphisms, Lorentz rotations,
supersymmetries and Maxwell transformations, respectively. For the
time being, the spin
connection is treated as an independent variable: later on we shall
go over to second
order formalism. It is useful to similarly
extend the operation of exterior differentiation, as already mentioned
in (\ref{unotre}).
The curvatures are extended to ghost-forms of ghost-form-number two,
that are the
sum of a (2,0)-piece (the original curvature) plus a (1,1)-term and a
(0,2)-term.
These extra-terms will be fixed by the rheonomic parametrizations. Let it be
\begin{eqnarray}
\hat R^a&=&R^a+\psi^a+\phi^a,\nonumber\\
\hat R^{ab}&=&R^{ab}+\chi^{ab}+\eta^{ab},\nonumber\\
\hat \rho_A&=&\rho_A+\xi_A+\zeta_A,\nonumber\\
\hat R^\otimes &=&R^\otimes+\psi+\phi.
\label{curvbrs}\end{eqnarray}
The same curvatures can be written by suitably extending
the definitions (\ref{curvatures})
(that is to say by replacing nonhatted quantities
with the corresponding hatted version).
For example,
\begin{equation}
\hat R^a=R^a+\psi^a+\phi^a=\hat d \hat V^a-\hat\omega^{ab}\wedge \hat V_b
-{i\over 2}\bar{\hat\psi}_A\wedge \gamma^a \hat\psi_A.
\end{equation}
After explicit substitution and separation of the various $(f,g)$-parts, one
can read, besides the definition of $R^a$ itself,
\begin{eqnarray}
sV^a&=&\psi^a-{\cal D}\varepsilon^a+\varepsilon^{ab}\wedge V_b+{i\over 2} (\bar
\psi_A\wedge\gamma^a c_A+\bar c_A\wedge\gamma^a \psi^A),\nonumber\\
s\varepsilon^a&=&\phi^a+\varepsilon^{ab}\wedge \varepsilon_b+{i\over 2}\bar
c_A\wedge \gamma^a c_A.
\end{eqnarray}
These are the BRST variations of $V^a$ and $\varepsilon^a$ (at least upon
fixing
$\psi^a$ and $\phi^a$). As $d^2=0$ and the extension to hatted quantities
preserves all the algebraic manipulations (as one can easily convince oneself),
we are guaranteed that this property is
extended to $\hat d^2=0$, that is to say
\begin{eqnarray}
d^2=0,\nonumber\\
ds+sd=0,\nonumber\\
s^2=0.
\label{nil}
\end{eqnarray}
In particular we are guaranteed to find a well defined BRST algebra ($s^2=0$).
By analysing the remaining curvatures in a similar way,
one gets the rest of the BRST algebra. We give only the results. The complete
BRST algebra is
\begin{eqnarray}
sV^a&=&\psi^a-{\cal D}\varepsilon^a+\varepsilon^{ab}\wedge V_b+i\bar c_A
\wedge\gamma^a\psi_A,\nonumber\\
s\omega^{ab}&=&\chi^{ab}-{\cal D}\varepsilon^{ab},\nonumber\\
s\varepsilon^a&=&\phi^a+\varepsilon^{ab}\wedge\varepsilon_b+{i\over 2}\bar c_A
\wedge \gamma^ac_A,\nonumber\\
s\varepsilon^{ab}&=&\eta^{ab}+{\varepsilon^a}_c\wedge\varepsilon^{cb},
\nonumber\\
s\psi_A&=&\xi_A-{\cal D}c_A+{1\over 2}\varepsilon^{ab}\sigma_{ab}\psi_A,
\nonumber\\
sc_A&=&\zeta_A+{1\over 2}\varepsilon^{ab}\sigma_{ab}c_A,\nonumber\\
sA&=&\psi-dc-2\epsilon_{AB} \bar c_A\wedge \psi_B,\nonumber\\
sc&=&\phi-\epsilon_{AB}\bar c_A\wedge c_B
\label{brs}
\end{eqnarray}
In a similar way one can also analyse the content of the hatted extensions of
the Bianchi identities (\ref{bianchi}).
One then finds two sets of variations:
i) the variations of the curvatures themselves, i.\ e.\
\begin{eqnarray}
sR^a&=&-{\cal D}\psi^a+\varepsilon^{ab}\wedge R_b-R^{ab}\wedge \varepsilon_b-
\chi^{ab}\wedge V_b +i\bar\psi_A\wedge \gamma^a \xi_A+i\bar c_A\wedge \gamma^a
\rho_A,\nonumber\\
sR^{ab}&=&-{\cal D}\chi^{ab}+{\varepsilon^a}_c\wedge R^{cb}-{R^a}_c
\wedge \varepsilon^{cb},\nonumber\\
s\rho_A&=&-{\cal D}\xi_A+{1\over 2}\varepsilon^{ab}\sigma_{ab}\rho_A-{1\over 2}
R^{ab}\sigma_{ab}c_A-{1\over 2}\chi^{ab}\sigma_{ab}\psi_A,\nonumber\\
sR^{\otimes}&=&-d\psi-2\epsilon_{AB}(\bar\psi_A\wedge \xi_B+
\bar c_A\wedge\rho_B),
\label{brscurv}
\end{eqnarray}
that are consistent with their definitions (\ref{curvatures}) and with
(\ref{brs});
ii) the variations of the {\sl free parameters} $\psi^a$, $\phi^a$,
$\chi^{ab}$, $\eta^{ab}$,
$\xi_A$, $\zeta_A$, $\psi$ and $\phi$,
\begin{eqnarray}
s\psi^a&=&-{\cal D}\phi^a+\varepsilon^{ab}\wedge\psi_b-\chi^{ab}\wedge
\varepsilon_b-\eta^{ab}\wedge V_b+i\bar\psi_A\wedge\gamma^a\zeta_A+
i\bar c_A\wedge\gamma^a\xi_A,\nonumber\\
s\phi^a&=&\varepsilon^{ab}\wedge\phi_b-\eta^{ab}\wedge\varepsilon_b
+i\bar c_A\wedge\gamma^a\zeta_A,\nonumber\\
s\chi^{ab}&=&-{\cal D}\eta^{ab}+\varepsilon^{ac}\wedge{\chi_c}^b-\chi^{ac}
\wedge {\varepsilon_c}^b,\nonumber\\
s\eta^{ab}&=&\varepsilon^{ac}\wedge{\eta_c}^b-\eta^{ac}\wedge
{\varepsilon_c}^b,
\nonumber\\
s\xi_A&=&-{\cal D}\zeta_A+{1\over 2}\varepsilon^{ab}\sigma_{ab}
\xi_A-{1\over 2}
\chi^{ab}\sigma_{ab}c_A-{1\over 2}\eta^{ab}\sigma_{ab}\psi_A,\nonumber\\
s\zeta_A&=&{1\over 2}\varepsilon^{ab}\sigma_{ab}\zeta_A-
{1\over 2}\eta^{ab}\sigma_{ab}c_A, \nonumber\\
s\psi&=&-d\phi-2\epsilon_{AB}(\bar c_A\wedge\xi_B+\bar\psi_A\wedge \zeta_B),
\nonumber\\
s\phi&=&-2\epsilon_{AB}\bar c_A\wedge \zeta_B.
\label{brsaux}
\end{eqnarray}
Eq.s (\ref{brsaux}) and (\ref {brscurv}) are the specialization to the case
of the BRST quantum algebra of N=2 supergravity of the last three
equations in (\ref{unouno}).

As a next step, we fix the free parameters ($\psi^a$, $\phi^a$, $\chi^{ab}$,
$\eta^{ab}$,
$\xi_A$, $\zeta_A$, $\psi$ and $\phi$) by means of the rheonomic conditions
\cite{beaulieu,fre}. These conditions state that the parametrizations of the
hatted curvatures are obtained by the old ones (see (\ref{rheo})) upon
substitution of the forms $V^a$ and $\psi_A$ (the basis of forms in superspace)
by their hatted quantities. For example, according to this prescription,
$\hat\rho_A$ is equal to
\begin{equation}
\hat\rho_A=\rho^A_{ab}\hat V^a\wedge \hat V^b+{1\over 2}i\gamma^a{\cal F}_{ab}
\hat\psi_B\wedge\hat V^b \epsilon_{AB}.
\end{equation}
After use of (\ref{hat}) and (\ref{curvbrs}) and separation of the various
$(f,g)$-parts, one
can read the definitions of $\xi_A$ and $\zeta_A$. In a similar way one
prodeeds
for the other curvatures and free parameters. We report here only the final
result, that is
\begin{eqnarray}
\psi^a&=&0,\nonumber\\
\phi^a&=&0,\nonumber\\
\chi^{ab}&=&2 {R^{ab}}_{cd} V^c \wedge \varepsilon^d+{\bar\theta^{ab}}_{A|c}
(c_A \wedge V^c+\psi_A \wedge \varepsilon^c)-\bar c_A {\cal F}^{ab}\psi_B
\epsilon_{AB},\nonumber\\
\eta^{ab}&=&{R^{ab}}_{cd}\varepsilon^c\wedge\varepsilon^d
+{\bar\theta^{ab}}_{A|c}
c_A \wedge \varepsilon^c-{1\over 2}\bar c_A {\cal F}^{ab}c_B\epsilon_{AB},
\nonumber\\
\xi_A &=&2\rho^A_{ab}\varepsilon^a\wedge V^b+{i\over 2}\gamma^a
{\cal F}_{ab}(c_B\wedge V^b+\psi_B\wedge\varepsilon^b)\epsilon_{AB},
\nonumber\\
\zeta_A &=&\rho^A_{ab}\varepsilon^a\wedge\varepsilon^b+{i\over 2}\gamma^a
{\cal F}_{ab}c_B\wedge\varepsilon^b\epsilon_{AB},\nonumber\\
\psi &=&2F_{ab}V^a\wedge\varepsilon^b,\nonumber\\
\phi &=&F_{ab}\varepsilon^a\wedge\varepsilon^b.
\label{freepar}
\end{eqnarray}
One can verify that (\ref{brsaux}) are consistent with (\ref{freepar}) on
shell.
This requires no further computational work than the one which is
required to prove that the rheonomic
parametrizations (\ref{rheo}) are consistent with the Bianchi identities
(\ref{bianchi}) \cite{fre} (the formal manipulations are the same).

By means of suitable redefinitions one can put formulas (\ref{brs}) in a more
familiar form, i.\ e.\ to write diffeomorphisms in terms of Lie
derivatives \cite{beaulieu}.
To this purpose, let $\varepsilon^\mu=\varepsilon^a V^\mu_a$, $V^\mu_a$ being
the inverse vierbein, so that $\varepsilon^a=i_{\varepsilon}V^a$, where $i$
denotes contraction. The Lie derivative ${\cal L}_\varepsilon$ is equal to
$i_\varepsilon d-d i_\varepsilon$, where the minus sign is due to the fact that
$\varepsilon$ is a ghost \cite{beaulieu}.
If we define ${\varepsilon^\prime}^{ab}=\varepsilon^{ab}-i_\varepsilon
\omega^{ab}$, $c^\prime_A=c_A-i_\varepsilon \psi_A$ and
$c^\prime=c-i_\varepsilon A$, then we get
\begin{eqnarray}
sV^a&=&{\cal L}_\varepsilon V^a+{\varepsilon^\prime}^{ab}\wedge V_b+i\bar
c^\prime_A\wedge\gamma^a\psi_A,\nonumber\\
s\psi_A&=&{\cal L}_\varepsilon\psi_A+{1\over 2}{\varepsilon^
\prime}^{ab}\sigma_{ab}\psi_A-{\cal D}c^\prime_A+{i\over 2}\gamma^a
{\cal F}_{ab}c^\prime_B\wedge V^b
\epsilon_{AB},\nonumber\\
sA&=&{\cal L}_\varepsilon A-dc^\prime-2\epsilon_{AB}
 \bar c^\prime_A\wedge \psi_B.
\label{brsprime}
\end{eqnarray}
We see that the variations of the main fields $V^a$, $\psi_A$ and $A$
are the sum
of diffeomorphisms, Lorentz rotations and supersymmetries, as it must be.

The last point reguards the possiblity of employing the second order formalism,
that is to say of expressing $\omega^{ab}$ in terms of the vierbein ($R^a=0$).
For consistency, we must also have $sR^a=0$, and this gives a condition on
$\chi^{ab}$. Consequently, we should expect to have a condition on
$s\chi^{ab}$, however $s\chi^{ab}$ turns out to be automatically consistent
with (\ref{brs}) and so it imposes no further constraint.

N=2 supergravity has an internal $SU(2)_I$ symmetry holding off-shell and an
internal $U(1)$ symmetry, which, however, holds only on shell
\cite{nieuwenhuizen}
(that is to say it is a symmetry of the equations of motion, but not of the
lagrangian). This $U(1)$ internal symmetry combines
chirality of the gravitinos
with duality of the graviphoton in the following way \cite{nieuwenhuizen}
\begin{eqnarray}
\delta \psi_A&=&i\alpha \gamma_5 \psi_A,\nonumber\\
\delta F_{ab}&=&-2i\alpha \tilde F_{ab}=\alpha \epsilon_{abcd}F^{cd}.
\label{chidua}
\end{eqnarray}
One easily verifies that the equations of motion derived from (\ref{lagra})
are all invariant under the chiral-dual transformations (\ref{chidua}).
The less trivial case is the one of the field equation coming
from the variation of the graviphoton $A$, that is
\begin{equation}
4i\epsilon_{AB}\bar\rho_A\wedge \gamma_5\psi_B-{\cal D}(F^{ab}V^c\wedge V^d)
\epsilon_{abcd}=0.
\end{equation}
Its variation under
(\ref{chidua}) is the last
of the Bianchi identities (\ref{bianchi}) and viceversa, thus proving
U(1) on shell
invariance (the remaining Bianchi identities are trivially invariant).

\section{Topological Gravity}
\label{topgrav}

In this section we discuss the gauge-free BRST algebra of topological gravity.
As already pointed out this BRST algebra involves only ghosts (and not
antighosts).
In the following sections we  show that this algebra stands to the algebra
of twisted topological gravity determined in Section
\ref{brsn=2} as the Beaulieu-Singer approach \cite{beaulieuYM} stands to  the
Witten approach \cite{witten}.
The procedure resembles the construction of the BRST quantum
version of supergravity,
but the difference is that, according to the discussion of Section I,
we impose no rheonomic parametrization. We
show that this prescription gives automatically a topological theory.
Similarly, (\ref{brs}) and (\ref{brsaux}),
without imposition of (\ref{freepar}), are the gauge-free BRST algebra of
topological N=2 supergravity.

Let us start from
the curvatures of the theory, that are given by (\ref{curvtop}).
Similarly with respect to before, we define hatted quantities
\begin{eqnarray}
\hat d&=&d+s,\nonumber\\
\hat V^a&=&V^a+\varepsilon^a,\nonumber\\
\hat\omega^{ab}&=&\omega^{ab}+\varepsilon^{ab},\nonumber\\
\hat R^a&=&R^a+\psi^a+\phi^a,\nonumber\\
\hat R^{ab}&=&R^{ab}+\chi^{ab}+\eta^{ab},
\end{eqnarray}
but now $\psi^a$, $\phi^a$ $\chi^{ab}$ and $\eta^{ab}$ remain
independent fields.
{}From the definitions of the curvatures (\ref{curvtop}),
extended to hatted expressions as before,
and the Bianchi identities
\begin{eqnarray}
{\cal D}R^a+{R^a}_b \wedge V^b=0,
\nonumber\\
{\cal D}R^{ab}=0,
\end{eqnarray}
also extended to hatted quantities, one obtains the BRST algebra
\begin{eqnarray}
sV^a&=&\psi^a-{\cal D}\varepsilon^a+\varepsilon^{ab}\wedge V_b,\nonumber\\
s\omega^{ab}&=&\chi^{ab}-{\cal D}\varepsilon^{ab},\nonumber\\
s\varepsilon^a&=&\phi^a+\varepsilon^{ab}\wedge\varepsilon_b,\nonumber\\
s\varepsilon^{ab}&=&\eta^{ab}+{\varepsilon^a}_c\wedge\varepsilon^{cb},
\nonumber\\
s\psi^a&=&-{\cal D}\phi^a+\varepsilon^{ab}\wedge\psi_b-\chi^{ab}\wedge
\varepsilon_b-\eta^{ab}\wedge V_b,\nonumber\\
s\phi^a&=&\varepsilon^{ab}\wedge\phi_b-\eta^{ab}\wedge\varepsilon_b,\nonumber\\
s\chi^{ab}&=&-{\cal D}\eta^{ab}+\varepsilon^{ac}\wedge{\chi_c}^b-\chi^{ac}
\wedge{\varepsilon_c}^b,\nonumber\\
s\eta^{ab}&=&\varepsilon^{ac}\wedge{\eta_c}^b-\eta^{ac}\wedge{\varepsilon_c}^b.
\label{brstop}
\end{eqnarray}
Once more this is the specialization to the case of
the Poincar\'e algebra of Eq.s (\ref{unouno}).
Of course, this algebra is also obtainable by reduction to N=0 of the
N=2 algebra of
Eq.s\ (\ref{brs}) and (\ref{brsaux}) (with no imposition of (\ref{freepar})).

We have used the same symbols as before, for similar,
but different, quantities. Whenever
necessary, we shall distinguish objects belonging to the BRST
algebra of N=2 supergravity (\ref{brs})
from those of the BRST algebra of topological gravity (\ref{brstop})
by an index,
which will be 2 in
the former case, 0 in the latter.
For example, ${\omega_2}^{ab}$ will be the superconnection (coming from
${R_2}^a=0$), while
${\omega_0}^{ab}$ will be the usual connection (coming from ${R_0}^a=0$).
The transformations (\ref{brs}) will be denoted by $s_2$,
the transformations (\ref{brstop}) by $s_0$. Similarly,
we shall write ${\psi_2}^a$ and
${\psi_0}^a$, ${\phi_2}^a$ and ${\phi_0}^a$, et cetera.

As before, we are guaranteed that $s^2=0$, but now
$s_0^2=0$ holds off-shell (it is the imposition of a rheonomic
parametrization holding
only on shell that forces $s_2^2=0$ to hold only on shell).
Let us analyse (\ref{brstop})
in more detail.
As we see, ${\psi_0}^a$ represents the topological ghost and the variation of
$V^a$ is equal to the topological variation ${\psi_0}^a$ plus diffeomorphisms
plus Lorentz rotations. ${\phi_0}^a$ and
${\eta_0}^{ab}$ are ghosts for ghost, the former corresponding to
diffeomorphisms,
the latter corresponding to Lorentz rotations.
As for ${\chi_0}^{ab}$,
in the second order formalism ($R^a=0$) the condition
$s_0{R_0}^a=0$ (which can be read from the first formula of (\ref{brscurv})
upon
reduction to N=0)
implies $\chi_0^{ab}\wedge V_b=-{\cal D}_0{\psi_0}^a-{R_0}^{ab}
\wedge\varepsilon_b$, which
can be solved
in the same manner as the condition defining $\omega_0^{ab}$
(i.\ e.\ ${\omega_0}^{ab}\wedge V_b=dV^a$). As noted in Section
\ref{brsn=2}, the
fact that ${\chi_0}^{ab}$ depends on the other fields does
not impose further constraints and
the BRST algebra is well defined. From now on we shall employ
the second order formalism.

The procedure here followed to determine a BRST algebra
for topological gravity does not
introduce any antighost. This is because we are not choosing
any particular gauge-fixing.
The topological twist, on the other hand, will give automatically
a preferred gauge-fixing for the
topological symmetry, as we shall see in the following section.

Now we describe the observables of the theory, which are related to the
Pontriagin ${\cal P}={R_0}^{ab}\wedge {R_0}_{ab}$ and Euler
characteristic classes ${\cal E}={R_0}^{ab}\wedge
{R_0}^{cd}\varepsilon_{abcd}$.
$s{\phi_0}^a$ and $s{\eta_0}^{ab}$ should be compared with the variation of
the ghost for ghost $\phi$ that appears in (\ref{unouno}),
$s\phi=-[c,\phi]$. As we see, the transformation of $\phi$ is nothing but
a gauge transformation and so all gauge invariants constructed from $\phi$
are BRST invariants and can lead to the descent equations that give the
observables of the theory \cite{witten,beaulieuYM}. In our case it is
${\eta_0}^{ab}$ that has a BRST variation which is only a gauge transformation
(Lorentz rotation). ${\eta_0}^{ab}$ is a $4\times 4$ antisymmetric matrix. Any
$4\times 4$ matrix has the four invariants ${\rm tr}[{\eta_0}]$,
${\rm tr}[{\eta_0}^2]$, ${\rm tr}[{\eta_0}^3]$ and ${\rm tr}[{\eta_0}^4]$.
In our case only ${\rm tr}[{\eta_0}^2]$ and ${\rm tr}[{\eta_0}^4]$ are
nonvanishing. What are the corresponding descent equations and to what
topological
invariants do they correspond? It will be soon proved that they correspond
to the
Pontriagin number and to the Euler number.

We start by noticing that
the proof that the form ${R_0}^{ab}\wedge {R_0}_{ab}$ is
closed works with hatted quantities,
exactly as with nonhatted ones:
\begin{equation}
\hat d({{\hat R}_0}{}^{ab}\wedge {{\hat R}_0}{}_{ab})=
-2\hat d{{\hat R}_0}{}^{ab}\wedge {{\hat R}_0}{}_{ba}=
-2({{\hat\omega}_0}{}^{ab}\wedge {{\hat R}_0}{}_{bc}\wedge
{{{\hat R}_0}{}^c}{}_a-
{{\hat R}_0}{}^{ab}\wedge{{\hat\omega}_0}{}_{bc}\wedge{{{\hat
R}_0}{}^c}{}_a)=0.
\end{equation}

We have used the hatted Bianchi identity $\hat{\cal D}\hat R_0{}^{ab}=0$.
After explicit substitution and separation of the various $(f,g)$-parts,
one can read the
descent equations
\begin{eqnarray}
s_0\hskip .1truecm {\rm tr}[\eta_0\wedge \eta_0]&=&0,\nonumber\\
s_0\hskip .1truecm {\rm tr}[\eta_0\wedge\chi_0+\chi_0\wedge\eta_0]
&=&-d\hskip .1truecm
{\rm tr}[\eta_0\wedge \eta_0],\nonumber\\
s_0\hskip .1truecm {\rm tr}[\eta_0\wedge R_0+\chi_0\wedge\chi_0
+R_0\wedge\eta_0]&=&-
d\hskip .1truecm {\rm tr}
[\eta_0\wedge\chi_0+\chi_0\wedge\eta_0],\nonumber\\
s_0\hskip .1truecm {\rm tr}[R_0\wedge
\chi_0+\chi_0\wedge R_0]&=&-d\hskip .1truecm {\rm tr}[\eta_0\wedge R_0
+\chi_0\wedge\chi_0+
R_0\wedge\eta_0],\nonumber\\
s_0\hskip .1truecm {\rm tr}[R_0\wedge R_0]&=&-d\hskip .1truecm {\rm tr}
[R_0\wedge
\chi_0+\chi_0\wedge R_0],\nonumber\\
0&=&-d\hskip .1truecm {\rm tr}[R_0\wedge R_0],
\label{descent}
\end{eqnarray}
where the trace refers to the Lorentz indices. So,
we have the following observables
\begin{eqnarray}
{\cal O}^{(0)}&=&{\rm tr}[\eta_0\wedge\eta_0],\nonumber\\
{\cal O}^{(1)}_\gamma&=&\int_\gamma{\rm tr}[\eta_0\wedge\chi_0+\chi_0\wedge
\eta_0],\nonumber\\
{\cal O}^{(2)}_S&=&\int_S{\rm tr}[\eta_0\wedge R_0+\chi_0\wedge\chi_0
+R_0\wedge\eta_0],\nonumber\\
{\cal O}^{(3)}_V&=&\int_V{\rm tr}[R_0\wedge \chi_0+\chi_0\wedge R_0],
\nonumber\\
{\cal O}^{(4)}_{\cal M}&=&\int_{\cal M}{\rm tr}[R_0\wedge R_0],
\label{observables}
\end{eqnarray}
where ${\cal M}$ is the four dimensional manifold where the
theory is defined, and $\gamma$,
$S$, and $V$ are generic one-, two- and three-dimensional
cycles on ${\cal M}$.
So we have proved that ${\rm tr}[{\eta_0}^2]$ corresponds to the
Pontriagin number.
In precisely the same way, one can deduce descent equations
and construct observables associated to the Euler form
${\cal E}={R_0}^{ab}\wedge {R_0}^{cd}\epsilon_{abcd}$. These observables
will be denoted by $\tilde{\cal O}^{(n)}$ and correspond to
${\rm tr}[{\eta_0}\wedge \tilde{\eta_0}]$. As ${\rm tr}[{\eta_0}^4]=
{1\over 16}({\rm tr}[{\eta_0}\wedge \tilde{\eta_0}])^2+{1\over 2}
({\rm tr}[{\eta_0}^2])^2$, we see that we have exhausted the two
invariants discussed before.

\section{Topological twist of N=2 supergravity}
\label{toptwist}

The topological twist of N=2 supergravity is performed in a
similar way as the topological twist
of Yang Mills  theories \cite{witten}. Nevertheless, some generalizations and
specifications are
needed. We identify the internal symmmetry group
$SU(2)_I$ with $SU(2)_R$, the
right handed part of the Lorentz group, that is to say we define a twisted
$SU(2)_R^\prime$ as the diagonal subgroup
of $SU(2)_R\otimes SU(2)_I$. Let us fix a bit of notation.
Every field will be classified, before
the twist, by
an expression like $^c(L,R,I)^g_f$, where $L$, $R$ and $I$
are the representation labels for $SU(2)_L$,
$SU(2)_R$ and $SU(2)_I$ respectively, $c$ is the $U(1)$ charge,
$g$ is the ghost number and $f$
is the form degree. Some fields (the graviphoton and the
corresponding ghosts) have not a well
defined $U(1)$ charge and so $c$ will be replaced by a dot
in these cases. After the twist, each
field will be denodet by $(L,R^\prime)^{g+c}_f$, where
$R^\prime=R\otimes I$. The new ghost number
is the sum of the old ghost number and the old $U(1)$ charge.
So, for some fields the new ghost
number is not defined off-shell, but only on shell. However, we
do not think this is a problem,
rather one of the new features of ghost number conservation in
topological theories. We note that
ghost number conservation has particular features even in twisted
Yang Mills theories \cite{witten},
because the chiral anomaly of the untwisted theory appears as a ghost
number anomaly in the twisted
version of the theory. In two dimensional topological theories, the
same phenomenon is represented
by the appearance of a charge at infinity after the twist \cite{2Dtwist}.
We think that the new
features of ghost number conservation that appear in twisted N=2
supergravity deserve further
investigation.

The fields are also characterized by a fermionic number, however
it will not play an important
role in the twisted theory. We shall explain this fact in the following
section.

In Table \ref{twist} we list the fields of N=2 supergravity and
their twisted counterparts. We
see that the twisted version of $\psi_A$
has a $({1\over 2},{1\over 2})^1_1$ component. This is
substantially the ghost of topological variations of
the vierbein (the exact identification will be given in the following section).
The components $(0,1)^{-1}_1$ and $(0,0)^{-1}_1$
become the corresponding antighosts.
The variation of the  $(0,1)^{-1}_1$ component, in particular,
gives the gauge-fixing of the
topological symmetry, precisely as in Yang-Mills theories. The
twisted version of $A$ represents
ghosts for ghosts and antighosts for ghosts. This is because the
tensor $F^{ab}$
has two components of $U(1)$ charge $\pm 2$, and so the twisted
version of $F^{ab}$ has two
components of ghost number $\pm 2$.

In Table \ref{twaux} we list the antighosts
and Lagrange multipliers of N=2 supergravity and their twisted counterparts.
$\bar\varepsilon^a$ and $\bar\varepsilon^{ab}$ are the antighosts of
diffeomorphisms
and Lorentz rotations, respectively; $\pi^a$ and $\pi^{ab}$ are the
corresponding
Lagrange multipliers; $\bar c_A^*$ are the antighosts of supersymmetries and
$P_A$ are their Lagrange multipliers; $\bar c$ and $P$ are the antighost and
Lagrange multiplier of the Maxwell gauge-symmetry.
In Table \ref{summary} we give a summary
of all the fields involved in the BRST quantum algebra of N=2 supergravity,
their twisted version and their meaning.

The explicit twist can be realized by interpreting the internal
indices $A$, $B$ as dotted
indices $\dot\alpha$,  $\dot\beta$.
Refer to Appendix \ref{notation} for the notation. The left handed and right
handed components
of $\psi_A$ are twisted as follows
\begin{eqnarray}
 {\psi_\alpha}_A&\rightarrow&\psi_{\alpha\dot A},\nonumber\\
 {\psi^{\dot\alpha}}_A &\rightarrow&\psi^{\dot\alpha\dot A},
\end{eqnarray}
while $\epsilon_{AB}\rightarrow\epsilon_{\dot A\dot B}=
-\epsilon^{\dot A\dot B}$. Let us now consider the
supersymmetry transformations (which can be read from (\ref{brsprime}) when
$\varepsilon^a=0$ and $\varepsilon^{ab}=0$)
\begin{eqnarray}
\delta V^a&=&i\bar c_A\wedge\gamma^a\psi_A,\nonumber\\
\delta\psi_A&=&-{\cal D}c_A+{i\over 2}\gamma^a
{\cal F}_{ab}c_B\wedge V^b\epsilon_{AB},\nonumber\\
\delta A&=&-2\epsilon_{AB} \bar c_A\wedge \psi_B.
\label{susy}
\end{eqnarray}
We now twist these transormations
and specialize the twisted version of the supersymmetry ghost
$c_A$ to its $(0,0)^0_0$-component.
This component is $C\equiv{c_{\dot\alpha}}^{\dot A}\delta_{\dot A}^
{\dot\alpha}$ (see Appendix
\ref{notation}). We set it equal to a constant and precisely $+i$,
for convenience.
The twisted version of ${\psi}_A$ consists of a $({1\over 2},
{1\over 2})^1_1$-component, which
will be denoted by ${\tilde\psi}^m=-{1\over 2}{\psi}_{\alpha\dot A}
(\bar\sigma^m)^{\dot A\alpha}$ (the ghost of topolgical symmetry),
a $(0,1)^{-1}_1$-component, which will be denoted by
${\tilde\psi}^{ab}={(\bar\sigma^{ab})^{\dot A}}_{\dot\alpha}{{\psi}
^{\dot \alpha}}_{\dot A}$ (the antighost corresponding to the gauge
breaking of the topological symmetry) and a
$(0,0)^{-1}_1$-component, denoted by $\tilde\psi={{\psi}_{\dot\alpha}}
^{\dot A}
\delta_{\dot A}^{\dot\beta}$
(see Table \ref{twist} and Appendix \ref{notation}).
The transformations (\ref{susy}) become
\begin{eqnarray}
\delta V^a&=&\tilde\psi^a,\nonumber\\
\delta \tilde\psi^a&=&{1\over 4}{F^-}^{ab}\wedge V_b,\nonumber\\
\delta \tilde\psi^{ab}&=&{i\over 4}{\omega^-}^{ab},\nonumber\\
\delta \tilde\psi &=&0, \nonumber\\
\delta A &=& i \tilde\psi.
\label{twsusy}
\end{eqnarray}
These transformations should be compared with those of topological Yang-Mills
theories, as found in Ref.\ \cite{witten}.
As we see, the topological gauge-fixing is the antiselfdual part
of the spin connection. Its
vanishing describes the gravitational instantons of the theory
of topological gravity that we
are studying.

The square of the transformation (\ref{twsusy}) is not zero, but it is a
Lorentz rotation with field-dependent
parameters ${1\over 4}{F^-}^{ab}$. This can
be immediately deduced from
the fact that $\delta^2 V^a=\delta \tilde\psi^a={1\over 4}
{F^-}^{ab}\wedge V_b$.
A phenomenon like the present one also
happens in topological Yang-Mills theories \cite{witten}.
It is only when dealing with
the complete BRST algebra \cite{beaulieu}
that the square of the transformations  is zero (at least on shell).

In \cite{beaulieu} we see that the complete BRST symmetry
of topological Yang-Mills theories
derives from a composition of the BRST symmetry of the
untwisted N=2 super Yang-Mills theory and
the $(0,0)^0_0$-component of the supersymmetric transformations.
We want to consider the analogue of this mechanism in twisted
topological gravity.
Here we have to deal with
the fact that now supersymmetry is a local
symmetry and nevertheless we expect that the twisted BRST
symmetry is in some sense a composition
of the twisted version of the transformations (\ref{brs})
and the transformations (\ref{twsusy}).
At the same time we need to be sure that the new BRST
symmetry closes on shell.
We cannot simply specialize the twisted version of supersymmetry
transformations to their  $(0,0)^0_0$-component, because this
would require to set some ghosts equal to zero, thus not
guaranteeing $s^2=0$.
A simple way to overcome
all this is to shift the $(0,0)^0_0$-component $C$ of
the twisted version of the ghosts
$c_A$ by a constant, namely $C\rightarrow C+i$. In such
a way the new BRST transformations
of the main
fields are the old ones plus the transformations (\ref{twsusy}),
as we would like, and
closure on shell is automatically assured.
The procedure of shifting $C$ will be called {\sl topological shift}.
The topological shift should be considered as a mere trick
to reach our purpose to define a
suitable new BRST symmetry and should not be reguarded as substantial.

The twisted-shifted BRST symmetry will be denoted by $s^\prime$.
It will not be explicitly
written down here; we only make some observations. The topological
twist and the topological shift make the new BRST
transformations appear as follows
\begin{eqnarray}
s^\prime V^a&=&\tilde\psi^a-d\varepsilon^a+\varepsilon^{ab}\wedge V_b
+\cdots,\nonumber\\
s^\prime \omega^{ab}&=&\chi^{ab}-d\varepsilon^{ab}+\cdots,\nonumber\\
s^\prime \varepsilon^{a}&=&C^a+\cdots,\nonumber\\
s^\prime \varepsilon^{ab}&=&-{1\over 4}{F^-}^{ab}+\cdots,\nonumber\\
s^\prime \eta^{ab}&=&0+\cdots,\nonumber\\
s^\prime\tilde\psi^a&=&-d C^a+{1\over 4}{F^-}^{ab}\wedge V_b+\cdots,
\nonumber\\
s^\prime \tilde\psi^{ab}&=&-dC^{ab}+{i\over 4}{\omega^-}^{ab}+\cdots,
\nonumber\\
s^\prime \tilde\psi &=&-dC +\cdots,\nonumber\\
s^\prime C^{a}&=&0+\cdots,\nonumber\\
s^\prime C^{ab}&=&{i\over 4}{\varepsilon^-}^{ab}+\cdots,\nonumber\\
s^\prime C&=&0+\cdots,\nonumber\\
s^\prime A &=& i \tilde\psi -dc+\cdots,\nonumber\\
s^\prime c &=&-{1\over 2}+iC+\cdots.
\label{topbrs}
\end{eqnarray}
where $\chi^{ab}\wedge V_b=-d\tilde\psi^a+\cdots$ and the dots refer
to interactions terms,
i.\ e.\ terms involving products of two or more fields.
$C^a=-{1\over 2}{c}_{\alpha\dot A}(\bar\sigma^a)^{\dot A\alpha}$,
${C}^{ab}={(\bar\sigma^{ab})^{\dot A}}_{\dot\alpha}{{c}^{\dot \alpha}}
_{\dot A}$ and
$C={c_{\dot\alpha}}^{\dot A}\delta_{\dot A}^{\dot\alpha}$
(see Table \ref{twist} and Appendix \ref{notation}). The BRST
transformation of the Maxwell
ghost $c$ contains a constant $-{1\over 2}$. This constant is
inessential and can be
suppressed. In fact $c$ appears only in $s^\prime A $ as $dc$
(not even the
dots contain $c$). Consequently, ${s^\prime}^2=0$ is assured even
if we write $s^\prime c =
iC+\cdots$. The transformations (\ref{topbrs})
will be useful for the computations of the next section; in particular,
note that, according to (\ref{freepar}), $\eta^{ab}=-{1\over 4}{F^-}^{ab}
+\cdots$, so
(\ref{topbrs}) shows that $s^\prime \varepsilon^{ab}=\eta^{ab}+\cdots$, in
agreement with (\ref{brstop}).

\section{Matching between twisted N=2 supergravity and topological gravity}
\label{matching}

In this section we give the correspondence between the transformations
$s^\prime$ (i.\ e.\
the topologically twisted
and topologically shifted version of $s_2$ (\ref{brs})) and the
transformations $s_0$
(\ref{brstop}), at least for what reguards the sector that not
includes antighosts.
First of all, let us compare
\begin{equation}
s_2 V^a=-{\cal D}_2\varepsilon^a+\varepsilon^{ab}\wedge V_b+i\bar c_A
\wedge\gamma^a\psi_A
\end{equation}
with
\begin{equation}
s_0 V^a=\psi_0^a-{\cal D}_0\varepsilon^a+\varepsilon^{ab}\wedge V_b.
\end{equation}
Let us put $\omega_2^{ab}=\omega_0^{ab}-A^{ab}$, where $A^{ab}$
is determined by the condition
$A^{ab}\wedge V_b={i\over 2}\bar \psi_A\wedge\gamma^a\psi_A$.
We can identify the two variations of $V^a$ (i.\ e.\ impose $s_2V^a
=s_0V^a$) if we put
\begin{equation}
{\psi_0}^a=i\bar c_A\wedge \gamma^a\psi_A-A^{ab}\wedge \varepsilon^b
={\tilde\psi}^a+\cdots.
\end{equation}
As we see, there is no need to make the topological twist and the
topological shift explicit.
By comparing
\begin{equation}
s_2\varepsilon^a=\varepsilon^{ab}\wedge\varepsilon_b+{i\over 2}\bar
c_A\wedge \gamma^ac_A
\end{equation}
with
\begin{equation}
s_0\varepsilon^a={\phi_0}^a+\varepsilon^{ab}\wedge\varepsilon_b,
\end{equation}
we deduce
\begin{equation}
{\phi_0}^a={i\over 2}\bar c_A\wedge\gamma^a c_A=C^a+\cdots.
\end{equation}
By comparing $s_2\varepsilon^{ab}$ and $s_0\varepsilon^{ab}$,
one deduces ${\eta_2}^{ab}=
{\eta_0}^{ab}$. After making these identifications, one can check
by a direct but tedious computation
that $s_2{\psi_0}^a$, $s_2{\phi_0}^a$, $s_2{\chi_0}^{ab}$
$s_2{\eta_0}^{ab}$ and $s_2{\omega_0}^{ab}$ automatically match with
the corresponding $s_0$-transformations.

Let us make a comment to the fact that the above identifications
involve bilinear terms
in the fields. This problem is promptly solved by the topological
shift that reduces the
above products of fields to a term linear in the fields plus
interactions. The presence of these interactions is the consequence of the
already noted fact that, since supersymmetry is local, we cannot specialize it
to its $(0,0)^0_0$ component.
Moreover, a
little insight shows that the appearance of the bilinear terms
is all but a problem.
In fact, we must remember that in the topological twist chirality
adds to ghost
number; since the commutation properties between $s$ and the fields
is regulated by
ghost-form number, it could happen, in general, that a field changes
its properties
of commutation with $s$ during the twist. This would be surely
dangerous, because it
is important to preserve all formal manipulations to guarantee
$s^2=0$ on shell.
Having to deal with bilinear terms, we are sure that the commutation
properties do
not change, since chirality is always even. An analogous observation
can be done
about fermion number; furthermore, since in the twisted theory
fermion number has no importance,
we can simply forget about it. By means of bilinear terms it is
also possible to
define the antighosts ${\psi_0}^{ab}$ and ${\psi_0}$, at least up to
interaction pieces. The bilinear terms corresponding to them are
$\bar c_A \wedge
\sigma^{ab}{1-\gamma_5 \over 2}\psi_B \epsilon_{AB}$ and $\bar
c_A \wedge \psi_B
\epsilon_{AB}$.

The above identifications permit to get explicitly the observables
of the twisted theory,
by simply taking the definitions (\ref{observables}), rewriting
them in the twisted
notation and shifting the gost $C$. All that is needed is ${\chi_0}^{ab}$
and ${\eta_0}^{ab}$,
which are given by
\begin{eqnarray}
{\eta_0}^{ab}&=&{R_2^{ab}}_{cd}\varepsilon^c\wedge\varepsilon^d+
{\bar\theta^{ab}}_{A|c}
c_A \wedge \varepsilon^c-{1\over 2}\bar c_A {\cal F}^{ab}c_B\epsilon_{AB},
\nonumber\\
{\chi_0}^{ab}\wedge V_b&=&-{\cal D}_0\psi_0^a-R_0^{ab}\wedge\varepsilon_b.
\end{eqnarray}

\section{The lagrangian of topological gravity}
\label{toplagra}

In this section we discuss the twisted-shifted version of
the lagrangian
(\ref{lagra})
of N=2 supergravity. In particular, we want to show that it can be
written as the BRST
variation of a gauge fermion $\Psi$. We shall be satisfied of a
gauge fermion $\Psi$
that reproduces the kinetic terms of the twisted-shifted N=2
supergravity lagrangian.
In fact any gauge fermion can be corrected by adding to it
interaction terms and indeed,
even if they are not explicitly written, they are in general
required in perturbation
theory (a similar remark is made in \cite{beaulieu}). In any
case, the main requirement
for a good gauge fermion is that it must remove all degeneracies
of the kinetic terms
and permit the definition of propagators, so we are justified
in concentrating our
attention on the kinetic terms, a restriction that simplifies
considerably the
computational effort.

First of all, we re-write the gravitational action in the
second order formalism in
a convenient way. As a matter of fact, one easily verifies that
\begin{eqnarray}
{\cal L}&=&R^{ab}\wedge V^c \wedge V^d \epsilon_{abcd}=\nonumber\\
&=&2\omega^{ab}\wedge {\omega^c}_e\wedge V^e \wedge V^d \epsilon_{abcd}+
\nonumber\\
&-&\omega^{ae}
\wedge{\omega_e}^b\wedge V^c \wedge V^d
\epsilon_{abcd}+d(\omega^{ab}\wedge V^c \wedge V^d \epsilon_{abcd}),
\label{lagrangian}
\end{eqnarray}
and that
\begin{eqnarray}
{\cal A}&\equiv & {\omega^-}^{ab}\wedge {\cal M}_{ab,cd}\wedge
{\omega^-}^{cd}\equiv\nonumber\\
&\equiv &{\omega^-}^{ab}\wedge {{\omega^-}^c}_e\wedge V^e \wedge V^d
\epsilon_{abcd}-{1\over 2}
{\omega^-}^{ae}\wedge{\omega^-_e}^b\wedge V^c \wedge V^d
\epsilon_{abcd}=\nonumber\\
&=&{\cal L}-d(\omega^{ab}\wedge V^c \wedge V^d \epsilon_{abcd}
-2i V^a\wedge dV_a).
\end{eqnarray}
In other words, we have written the gravitational lagrangian as quadratic
in the antiselfdual part of the spin connection, which is our gauge-fixing,
plus
a total derivative.
${\cal M}_{ab,cd}$ is a two form and is independent from
derivatives of the vierbein. This way of expressing the gravitational
lagrangian (up to a topological term) should be compared with the
expression $-{1\over 4} {\rm tr}[F^-_{\mu\nu}{F^-}^{\mu\nu}]$ for the
lagrangian
of Yang-Mills theories, which is the square of the gauge-fixing of topological
Yang-Mills theories
\cite{beaulieuYM}.
Making space-time components explicit, we can write
\begin{equation}
{\cal A}\equiv {\omega_\mu^-}^{ab}\wedge {\cal M_{\nu,\rho}}_{ab,cd}\wedge
{\omega_\sigma^-}^{cd}
\epsilon^{\mu\nu\rho\sigma}.
\end{equation}
${{\cal M}_{\nu,\rho}}_{ab,cd}$ is a matrix which is
antysimmetric and antiselfdual
in $ab$ and $cd$. One can easily verify that there
exist only two such matrices in flat space,
namely the identity
${{\cal I}_{\nu,\rho}}_{ab,cd}={1\over 4}\eta_{\nu\rho}
(\eta_{ac}\eta_{bd}-\eta_{ad}
\eta_{bc}-i\epsilon_{abcd})$,
where $\eta_{ab}={\rm diag}(1,-1,-1,-1)$, and ${{\cal M}
_{\nu,\rho}}_{ab,cd}$ itself.
Furthemore,  ${\cal M}$ is invertible (one proves that
${\cal M}^2$ is not proportional
to ${\cal M}$, so it must be a nontrivial linear combination
of ${\cal M}$ and ${\cal I}$).

First of all, we introduce a Lagrange multiplier $B^{ab}$
for the topological symmetry (a one form, antisymmetric and
antiselfdual in $ab$), such that $s^\prime \psi^{ab}=B^{ab}$
and $s^\prime B^{ab}=0$. In this section we omit the subscript
$0$ in ${\psi_0}^a$, ${\psi_0}^{ab}$ and ${\psi_0}$.
By comparing with the old expression for $s^\prime \psi^{ab}$,
(\ref{topbrs}), we see
that the linear part of the gauge-fixing term is ${i\over 4}
{\omega^-}^{ab}-dC^{ab}$,
that is to say there is a ghost term besides the expected term
${i\over 4}{\omega^-}^{ab}$.
We shall explain in a short time the reason for this presence.
In any case, we expect
that the gauge-fermion $\Psi$ contains a term
\begin{equation}
\Psi_1=8i(2i B^{ab}+{\omega^-}^{ab}+4i dC^{ab})\wedge
{\cal M}_{ab,cd}\wedge
\psi^{cd}.
\label{gf}
\end{equation}
Indeed, the BRST variation of $\Psi_1$, i.\ e.\ $s^\prime \Psi_1$,
contains a term
\begin{equation}
-8i(2i B^{ab}+{\omega^-}^{ab}+4i dC^{ab})\wedge {\cal M}_{ab,cd}\wedge
B^{ab},
\end{equation}
which, upon integration over $B^{ab}$ gives
\begin{equation}
({\omega^-}^{ab}+4i dC^{ab})\wedge {\cal M}_{ab,cd}\wedge
({\omega^-}^{cd}+4i dC^{cd}).
\end{equation}
So, the gravitational lagrangian (\ref{lagrangian}) is correctly
reproduced (at least
up to a topological term), but there are two more terms,
namely $8i dC^{ab}\wedge
{\cal M}_{ab,cd}\wedge{\omega^-}^{cd}$ and $-16 dC^{ab}\wedge
{\cal M}_{ab,cd}
\wedge dC^{cd}$. The first one is zero (or better, it is a total derivative),
because
\begin{equation}
d( {\cal M}_{ab,cd}\wedge{\omega^-}^{cd})\equiv 0,
\label{redundancy}
\end{equation}
as can be promptly
checked. The second term looks like, at first sight, the
kinetic lagrangian for
the ghosts $C^{ab}$, however this is not true, because the kinetic part of
$-16 dC^{ab}
\wedge {\cal M}_{ab,cd}\wedge dC^{cd}$ turns out to be zero. From
these remarks
we can deduce two considerations:
i) the gauge-fixing ${\omega^-}^{ab}=0$ is redundant, because
the one forms ${\omega^-}^{ab}$ are not independent, but are related
by the condition
(\ref{redundancy}), which holds identically, without
imposition of ${\omega^-}^{ab}=0$;
ii) the ghost $C^{ab}$ is an extraghost, i.\ e.\ a ghost the presence of
which is due to a redundancy of the gauge-fixing;
of course, it is associated to the redundancy (\ref{redundancy}) of the
topological gauge-fixing
conditions ${\omega^-}^{ab}=0$. A good treatment of such nontrivial ghosts of
vanishing ghost number can be found in Ref.\ \cite{batalin}, where
the case of the
antisymmetric tensor, call it $B_{\mu\nu}$, is explicitly exhibited.
In that case the BRST variation of
the antighost $\bar C_\mu$ ($\delta \bar C_\mu=\partial^\nu B_{\mu\nu}-
\partial_\mu c_1$) contains,
as well as the expected gauge-fixing term, $\partial^\nu B_{\mu\nu}$,
a term involving the extraghost $c_1$ and
giving information about the redundancy (which is $\partial_\mu
\partial_\nu B^{\mu\nu}\equiv 0$), precisely as it
happens in our case. However, in the simple example of the
antisymmetric tensor $B_{\mu\nu}$ the analogous term of
$-16 dC^{ab}\wedge {\cal M}_{ab,cd}\wedge dC^{cd}$ does give the
kinetic lagrangian
of the extraghost $c_1$ and so there is no further problem. In our case,
instead, this
does not happen (the reason is the richness of symmetries of our
theory, in particular
local supersymmetry). Since, as previously noted, only one matrix
with the properties
of ${\cal M}$ exists besides ${\cal M}$ itself, that is the
identity ${\cal I}$,
there is little to do: to give a kinetic term to $C^{ab}$,
it is necessary to
have a further extraghost, say ${C^*}^{ab}$ (of ghost
number zero, antiselfdual
in $ab$) and a kinetic term $d{C^*}^{ab}\wedge
{\cal I}_{ab,cd}\wedge dC^{cd}$,
that is to say ${C^*}^{ab} \Box {C}_{ab}$ plus interactions.
However, since we are only reinterpreting
a teory and we cannot construct it by hand, such a field
and such a kinetic term must
already be present. In particular ${C^*}^{ab}$ can only
come from the twist of the antighost $\bar c^*_A$ of N=2 local
supersymmetry and as a matter of fact, Table \ref{twaux}
shows that such a
field is indeed present and it is precisely the $(0,1)^{0}_0$-component
of the twisted version of $\bar c^*_A$.
So, all we have to do is
to check that the gauge-fermion
that breaks supersymmetry, say $\Psi_S$,
gives the correct kinetic lagrangian for $C^{ab}$ and ${C^*}^{ab}$.
Let us choose the most common
expression for $\Psi_S$, i.\ e.\
\begin{equation}
\Psi_S\equiv \bar c^*_A D\!\!\!\!\slash (\gamma^a V^\mu_a
{\psi_\mu}_A+\alpha P_A),
\end{equation}
where $P_A$ is the lagrange multiplier of
supersymmetries ($s^\prime c^*_A =P_A$, $s^\prime P_A=0$)
and $\alpha$ is a constant that is
usually determined in order to conveniently simplify the kinetic
term of gravitinos
($\alpha$ will
be of no importance for our purposes). In any case, the
BRST variation of $\Psi_S$
contains a term of the kind
\begin{equation}
\bar c^*_A \partial\!\!\!\slash \gamma^a V^\mu_a \partial_\mu c_A.
\end{equation}
As we expect, after the twist, the quadratic term in ${C^*}^{ab}-C^{cd}$ is
${C^*}^{ab} \Box {C}_{ab}$.

Let us now come back to the analysis of the BRST variation
of the gauge-fermion
$\Psi_1$, (\ref{gf}). The terms
$-8i(2i B^{ab}+{\omega^-}^{ab}+4i dC^{ab})\wedge
s^\prime({\cal M}_{ab,cd})\wedge
\psi^{cd}$ are only interaction terms and so we
discard them. Then there are the terms
$8is^\prime({\omega^-}^{ab}+4i dC^{ab})\wedge {\cal M}_{ab,cd}\wedge
\psi^{cd}$.
By looking at (\ref{topbrs}), one sees that the
term with $dC^{ab}$
in $\Psi_1$ is required in order to restore
invariance under Lorentz rotations
(i.\ e.\ in order to avoid kinetic terms like
$d{\varepsilon^-}^{ab}\wedge
{\cal M}_{ab,cd}\wedge\psi^{cd}$). So, the only
kinetic term coming from $\Psi_1$
that remains to be discussed is $8i{\chi^-}^{ab}
\wedge {\cal M}_{ab,cd}\wedge\psi^{cd}$.
This term reproduces the twisted version of the
Rarita-Schwinger action, precisely
the part that contains ${\psi}^a$ and ${\psi}^{ab}$,
which turns out to be
\begin{equation}
-16  d\psi^a \wedge\psi_{ab}\wedge V^b.
\end{equation}
The remaining piece  of the Rarita-Schwinger action, namely
\begin{equation}
-8 d\psi^a\wedge \psi\wedge V_a,
\end{equation}
can be retrieved by means of a further piece $\Psi_2$
to be added to the gauge fermion
$\Psi_1$. $\Psi_2$ must also give account of the kinetic term of the
graviphoton and turns out to be (remember $\eta^{ab}=-{1\over 4}{F^-}^{ab}
+\cdots$ and $R^\otimes=dA+\cdots$)
\begin{equation}
\Psi_2=8iR^\otimes\wedge \psi^a\wedge V_a+{2\over 3}\eta^{ab}\varepsilon_{ab}
V^i\wedge V^j\wedge V^k\wedge V^l \varepsilon_{ijkl}.
\end{equation}

Summarizing, the total gauge-fermion is
\begin{eqnarray}
\Psi&=&8i(2i B^{ab}+{\omega^-}^{ab}+4i dC^{ab})\wedge
{\cal M}_{ab,cd}\wedge
\psi^{cd}+\nonumber\\
&+&8iR^\otimes\wedge \psi^a\wedge V_a+{2\over 3}\eta^{ab}\varepsilon_{ab}
V^i\wedge V^j\wedge V^k\wedge V^l \varepsilon_{ijkl},
\end{eqnarray}
plus the usual terms that break diffeomorphisms, Lorentz
rotations, supersymmetries
(this one being in part already discussed) and Maxwell gauge-symmetry.

\section{Conclusions and outlook}
\label{conclusions}

Having shown that twisted N=2 supergravity is a formulation of D=4
topological gravity
it remains to be seen which of the N=2 supergravity correlators are
topological
and
how they are accordingly calculated using some version of intersection
theory on instanton
moduli-space. We postpone this investigation to a future publication.
The same we do
with the other obvious problem, namely the topological twisting of the
hypermultiplets
and their coupling to topological gravity. It is by now clear that,
notwithstanding several
quite important subtleties, there is a completely parallelel development of the
topological twisting programme in two and four dimensions and we think
it worth to
explore all its consequences.

\appendix{Notation and conventions}
\label{notation}

In this appendix we give the notation for spinor algebra.
The algebra of $\gamma$-matrices is
represented by
\begin{equation}
\gamma^m=\left(\begin{array}{cc}
 0 &   \sigma^m\\
  \bar\sigma^m &   0
\end{array}\right),
\end{equation}
where
\begin{equation}
\sigma^0=\left( \begin{array}{cc} -1  &  0\\
                 0 &  -1   \end{array} \right),  \hskip .3truecm
\sigma^1=\left( \begin{array}{cc}  0  &  1\\
                1  &  0    \end{array} \right),  \hskip .3truecm
\sigma^2=\left( \begin{array}{cc}  0  &  -i\\
                i  &  0    \end{array} \right),  \hskip .3truecm
\sigma^3=\left( \begin{array}{cc}  1  &  0\\
                0  & -1    \end{array} \right),  \hskip .3truecm
\end{equation}
\begin{equation}
(\bar\sigma^m)^{\dot \alpha\alpha}=\epsilon^{\dot\alpha\dot\beta}
\epsilon^{
\alpha\beta}(\sigma^m)_{\beta\dot\beta},
\end{equation}
and
\begin{equation}
\epsilon^{12}=\epsilon_{21}=1, \hskip .3truecm
\epsilon_{12}=\epsilon^{21}=-1.
\end{equation}
A Lorentz vector $v^m$ is represented by
\begin{equation}
v_{\alpha\dot\alpha}=(\sigma^m)_{\alpha\dot\alpha}v_m,
\end{equation}
and the inverse formula is
\begin{equation}
v^m=-{1\over 2}v_{\alpha\dot\alpha}(\bar\sigma^m){}^{\dot\alpha\alpha}.
\end{equation}
An antisymmetric tensor $F^{ab}$ is represented by
\begin{equation}
F^{ab}={1\over 2}({F^+}{}^{ab}+{F^-}{}^{ab})={1\over 2}({f_\beta^+}{}^\alpha
{(\sigma^{ab})_\alpha}{}^\beta+
{{f^-}^{\dot\beta}}_{\dot\alpha}{(\bar\sigma^{ab}){}^{\dot\alpha}}
_{\dot\beta}),
\end{equation}
where
\begin{eqnarray}
{F^\pm}^{ab}&=&F^{ab}\pm{i\over 2}\varepsilon^{abcd}F_{cd},\nonumber\\
{(\sigma^{ab})_\alpha}{}^\beta&=&{1\over 4}(\sigma^a_{\alpha\dot\alpha}
{\bar\sigma^b}{}^{\dot\alpha\beta}-
\sigma^b_{\alpha\dot\alpha}{\bar\sigma^a}{}^{\dot\alpha\beta}),
\nonumber\\
{(\bar\sigma^{ab}){}^{\dot\alpha}}_{\dot\beta}&=&{1\over 4}({\bar\sigma^a}
{}^{\dot\alpha\alpha}
\sigma^b_{\alpha\dot\beta}-
{\bar\sigma^b}{}^{\dot\alpha\alpha}\sigma^a_{\alpha\dot\beta})
\end{eqnarray}
A generic spinor $\psi_A$ is written as
\begin{equation}
\psi_A=\left(\begin{array}{c}  {\psi_\alpha}{}_A\\
                  {\psi^{\dot\alpha}}_A \end{array} \right),
\end{equation}
while
\begin{equation}
\bar\psi_A=( {\psi^\alpha}_A,\hskip .1truecm \psi_{\dot\alpha}{}_A),
\end{equation}
and indices are raised and lowered by means of $\epsilon^
{\alpha\beta}$ and
$\epsilon^{\dot\alpha\dot\beta}$.
For other details, see \cite{bagger}.

\begin{table}
\caption{Topological twist}
\begin{tabular}{lcc}
Field  &         Before the twist   &     After the twist\\
\tableline
$V^a$   &         $^0({1\over 2},{1\over 2},0)^0_1 $   &
 $({1\over 2},{1\over 2})^0_1$\\
$\varepsilon^a$   &   $^0({1\over 2},{1\over 2},0)^1_0$ &
     $({1\over 2},{1\over 2})^1_0$\\
$\varepsilon^{ab}$   & $^0(1,0,0)^1_0\oplus
 \hskip .1truecm^0(0,1,0)^1_0$&
$(1,0)^1_0\oplus (0,1)^1_0$\\
$\psi_A$  &  $^1({1\over 2},0,{1\over 2})^0_1
 \oplus \hskip .1truecm ^{-1}(0,{1\over 2},{1\over 2})^0_1$      &
 $({1\over 2},{1\over 2})^1_1\oplus (0,1)^{-1}_1\oplus (0,0)^{-1}_1$ \\
$c_A$ &   $^1({1\over 2},0,{1\over 2})^1_0\oplus
 \hskip .1truecm ^{-1}(0,{1\over 2},{1\over 2})^1_0$      &
 $({1\over 2},{1\over 2})^2_0\oplus (0,1)^{0}_0\oplus (0,0)^{0}_0$ \\
$A$ &   $^.(0,0,0)^0_1$  &   $(0,0)^._1$\\
$c$ &    $^.(0,0,0)^1_0$ &    $(0,0)^._0$
\end{tabular}
\label{twist}
\end{table}

\begin{table}
\caption{Twist of antighosts and Lagrange multipliers}
\begin{tabular}{lcc}
Field  &         Before the twist   &     After the twist \\
\tableline
$\bar\varepsilon^a$   &   ${}^0({1\over 2},{1\over 2},0)^{-1}_0$   &
               $({1\over 2},{1\over 2})^{-1}_0$\\
$\bar\varepsilon^{ab}$ &   ${}^0(1,0,0)^{-1}_0\oplus
\hskip .1truecm^0(0,1,0)
 ^{-1}_0$   &  $(1,0)^{-1}_0\oplus (0,1)^{-1}_0$\\
$\bar c^*_A$ &  ${}^{-1}({1\over 2},0,{1\over 2})^{-1}_0\oplus
 \hskip .1truecm ^{1}(0,{1\over 2},{1\over 2})^{-1}_0$  &
 $({1\over 2},{1\over 2})^{-2}_0\oplus (0,1)^{0}_0\oplus (0,0)^{0}_0$ \\
$\bar c$ &    ${}^.(0,0,0)^{-1}_0$ &    $(0,0)^._0$\\
$\pi^a$   &   ${}^0({1\over 2},{1\over 2},0)^{0}_0$  &
 $({1\over 2},{1\over 2})^{0}_0$\\
$\pi^{ab}$ &   ${}^0(1,0,0)^{-1}_0\oplus \hskip .1truecm^0(0,1,0)^{-1}_0$  &
 $(1,0)^{0}_0\oplus (0,1)^{-1}_0$\\
$P_A$ &   ${}^{-1}({1\over 2},0,{1\over 2})^{0}_0\oplus \hskip .1truecm ^{1}
(0,{1\over 2},{1\over 2})^{0}_0$   &
 $({1\over 2},{1\over 2})^{-1}_0\oplus (0,1)^{1}_0\oplus (0,0)^{1}_0$ \\
$P$  &   ${}^.(0,0,0)^0_0$  &   $(0,0)^._0$
\end{tabular}
\label{twaux}
\end{table}

\begin{table}
\caption{Summary of the fields of N=2 supergravity and their twisted versions}
\begin{tabular}{lcclcc}
\tiny{Field} &  \tiny{Meaning}   & \tiny{Classification} &
\tiny{Tw.-field} & \tiny{Twisted meaning} & \tiny{Tw.-classification}\\
\tableline
\scriptsize{$V^a$}   & \scriptsize{vierbein} &
\scriptsize{$^0({1\over 2},{1\over 2},0)^0_1 $}
   &  \scriptsize{$V^a$} & \scriptsize{vierbein} &
 \scriptsize{$({1\over 2},{1\over 2})^0_1$}\\
\tableline
{}&{}&{}& \scriptsize{$\psi^a$} &\scriptsize{topological ghost}&
\scriptsize{$({1\over 2},{1\over 2})^1_1$}\\
\scriptsize{$\psi_A$}&\scriptsize{gravitinos}
& \scriptsize{$^1({1\over 2},0,{1\over 2})^0_1
 \oplus \hskip .1truecm ^{-1}(0,{1\over 2},{1\over 2})^0_1$}  &\scriptsize{
$\psi^{ab}$}&
\scriptsize{topological antighost} & \scriptsize{$(0,1)^{-1}_1$} \\
{}&{}&{}& \scriptsize{$\psi$} & \scriptsize{antighost} &
\scriptsize{$(0,0)^{-1}_1$} \\
\tableline
\scriptsize{$A$} & \scriptsize{graviphoton} &
\scriptsize{$^.(0,0,0)^0_1$}& \scriptsize{$A$} &
\scriptsize{ghosts for ghosts}&   \scriptsize{$(0,0)^._1$}\\
\tableline
\scriptsize{$\varepsilon^a$}   & \scriptsize{ghost} &
\scriptsize{$^0({1\over 2},{1\over 2},0)^1_0$} & \scriptsize{$\varepsilon^a$}&
\scriptsize{ghost} & \scriptsize{$({1\over 2},{1\over 2})^1_0$}\\
\scriptsize{$\pi^a$}   &\scriptsize{L.\ multiplier}&
\scriptsize{${}^0({1\over 2},{1\over 2},0)^{0}_0$}  &  \scriptsize{$\pi^a$}   &
\scriptsize{L.\ multiplier}&
\scriptsize{ $({1\over 2},{1\over 2})^{0}_0$}\\
\scriptsize{$\bar\varepsilon^a$} &\scriptsize{antighost}  &
\scriptsize{${}^0({1\over 2},{1\over 2},0)^{-1}_0$ }  &
\scriptsize{$\bar\varepsilon^a$} &\scriptsize{antighost} &
\scriptsize{$({1\over 2},{1\over 2})^{-1}_0$}\\
\tableline
\scriptsize{$\varepsilon^{ab}$} &\scriptsize{ghost}  &
\scriptsize{$^0(1,0,0)^1_0\oplus
 \hskip .1truecm^0(0,1,0)^1_0$}&\scriptsize{    $\varepsilon^{ab}$}
&\scriptsize{ghost}  &
\scriptsize{$(1,0)^1_0\oplus (0,1)^1_0$}\\
\scriptsize{$\pi^{ab}$} & \scriptsize{L.\ multiplier}&
\scriptsize{${}^0(1,0,0)^{-1}_0\oplus \hskip .1truecm^0(0,1,0)^{-1}_0$}  &
\scriptsize{$\pi^{ab}$} & \scriptsize{L.\ multiplier}&
\scriptsize{$(1,0)^{-1}_0
\oplus (0,1)^{-1}_0$}\\
\scriptsize{$\bar\varepsilon^{ab}$} & \scriptsize{antighost} & \scriptsize{
${}^0(1,0,0)^{-1}_0\oplus
\hskip .1truecm^0(0,1,0)
 ^{-1}_0$}  &\scriptsize{$\bar\varepsilon^{ab}$} & \scriptsize{antighost} &
\scriptsize{$(1,0)^{-1}_0\oplus (0,1)^{-1}_0$}\\
\tableline
{}&{}&{}&\multicolumn{3}{r} {\it (continues on next page)}\\
\end{tabular}
\label{summary}
\end{table}

\begin{tabular}{lcclcc}
\tableline
\tableline
\multicolumn{3}{l} {\it (continues from previous page)}&{}&{}&{}\\
\tableline
{}&{}&{}&\scriptsize{$C^a$}&\scriptsize{ghost for ghost}&\scriptsize{
$({1\over 2},{1\over 2})^2_0$}\\
\scriptsize{$c_A$} & \scriptsize{ghost}&  \scriptsize{$^1({1\over 2},0,
{1\over 2})^1_0\oplus
 \hskip .1truecm ^{-1}(0,{1\over 2},{1\over 2})^1_0$}&\scriptsize{
$C^{ab}$}&\scriptsize{
extraghost}&
\scriptsize{ $(0,1)^{0}_0$ } \\
{}&{}&{}&\scriptsize{$C$}&\scriptsize{extraghost}&\scriptsize{
$(0,0)^0_0$}\\
\tableline
{}&{}&{}&\scriptsize{$P^a$}&\scriptsize{antighost}&\scriptsize{
$({1\over 2},{1\over 2})^{-1}_0$}\\
\scriptsize{$P_A$} &\scriptsize{L.\ multiplier}&
\scriptsize{${}^{-1}({1\over 2},0,{1\over 2})^{0}_0\oplus
\hskip .1truecm ^{1}
(0,{1\over 2},{1\over 2})^{0}_0$}   & \scriptsize{ $P^{ab}$}&
\scriptsize{ghost}&
\scriptsize{ $(0,1)^{1}_0$}\\
{}&{}&{}&\scriptsize{$P^\prime$}&\scriptsize{ghost}&
\scriptsize{$(0,0)^{1}_0$} \\
\tableline
{}&{}&{}&\scriptsize{${C^*}^a$}&\scriptsize{antighost for ghost}&
\scriptsize{$({1\over 2},{1\over 2})^{-2}_0$}\\
\scriptsize{$\bar c^*_A$} & \scriptsize{antighost}&
\scriptsize{${}^{-1}({1\over 2},0,{1\over 2})^{-1}_0\oplus
 \hskip .1truecm ^{1}(0,{1\over 2},{1\over 2})^{-1}_0$}
&\scriptsize{${C^*}^{ab}$}&
\scriptsize{extraghost}&
\scriptsize{ $(0,1)^{0}_0$}\\
 {}&{}&{}&\scriptsize{$C^*$}&\scriptsize{extraghost}&\scriptsize{
$(0,0)^{0}_0$} \\
\tableline
\scriptsize{$c$} &\scriptsize{ghost} &  \scriptsize{
$^.(0,0,0)^1_0$} &\scriptsize{$c$}&\scriptsize{ghost}&
\scriptsize{$(0,0)^._0$}\\
\scriptsize{$P$}  &\scriptsize{L.\ multiplier}&
\scriptsize{${}^.(0,0,0)^0_0$}  &\scriptsize{ $P$}&
\scriptsize{ghost} & \scriptsize{$(0,0)^._0$}\\
\scriptsize{$\bar c$} & \scriptsize{antighost}&
\scriptsize{  ${}^.(0,0,0)^{-1}_0$}&
\scriptsize{$\bar c$}&
\scriptsize{ghost}  &\scriptsize{ $(0,0)^._0$}\\
\tableline
\tableline
\end{tabular}

\end{document}